# STRUCTURE AND DYNAMICS OF THE 13/14 NOVEMBER 2012 ECLIPSE WHITE-LIGHT CORONA

J. M. Pasachoff,[1,9] V. Rušin,[2] M. Saniga,[3] B. A. Babcock,[4] M. Lu,[4] A. B. Davis,[4] R. Dantowitz,[5] P. Gaintatzis,[6] J. H. Seiradakis,[6] A. Voulgaris,[6] D. B. Seaton,[7] **K. Shiota[8]**

[1] Williams College—Hopkins Observatory, Williamstown, MA 01267-2565; eclipse@williams.edu
[2] Astronomical Institute, Slovak Academy of Sciences, 059 60 Tatranská Lomnica, Slovakia; vrusin@ta3.sk
[3] Institute for Discrete Mathematics and Geometry, Vienna University of Technology, Wiedner Hauptstrasse 8-10, A-1040 Vienna, Austria; metod.saniga@tuwien.ac.at
[4] Astronomy Department, Williams College, Williamstown, MA 01267-2565; bryce.a.babcock@williams.edu, muzhoulu@gmail.com, allen.b.davis@yale.edu
[5] Clay Center Observatory, Dexter Southfield School, Brookline, MA; rondantowitz@gmail.com
[6] Section of Astrophysics, Astronomy and Mechanics, Aristotle University of Thessaloniki, Thessaloniki, Greece; paulgai@yahoo.gr
[7] SIDC-Royal Observatory of Belgium, Avenue Circulaire 3, 1180 Brussels, Belgium; dseaton@oma.be
[8] Solar Eclipse Information Center, Chofu, Tokyo, Japan; salt-star@mbj.nifty.com
[9] Caltech, Pasadena, CA 91125

*Subject headings:* Eclipses — Sun: corona — Sun: coronal mass ejections (CMEs) — Sun: magnetic topology — Sun: chromosphere — Sun: UV radiation

## ABSTRACT

Continuing our series of observations of the motion and dynamics of the solar corona over the solar-activity cycle, we observed the corona from sites in Queensland, Australia, during the 13 (UT)/14 (local time) November 2012 total solar eclipse. The corona took the low-ellipticity shape typical of solar maximum (flattening index $\varepsilon = 0.01$), showing a change from the composite coronal images that we had observed and analyzed in this journal and elsewhere for the 2006, 2008, 2009, and 2010 eclipses. After crossing the northeast Australian coast, the rest of the path of totality was over the ocean, so further totality was seen only by shipborne observers. Our results include measurements of velocities of a coronal mass ejection; during the 36 minutes of passage from the Queensland coast to a ship north of New Zealand, we find a speed of 413 km s$^{-1}$, and we analyze its dynamics. We discuss the shapes and positions of several types of coronal features seen on our higher-resolution composite Queensland images of the solar corona, including, many helmet streamers, very faint bright and dark loops at the base of helmet streamers, voids and radially oriented thin streamers. We compare our eclipse observations with a hairy-ball model of the magnetic field, confirming the validity of the prediction, and we relate the eclipse phenomenology seen with the near-simultaneous images from the Atmospheric Imaging Assembly on the NASA's *Solar Dynamics Observatory* (*SDO*/AIA), the Extreme Ultraviolet Imager on NASA's *Solar Terrestrial Relations Observatory* (*STEREO*/EUVI), the Sun Watcher using Active Pixel System Detector and Image Processing on ESA's *PRoject for OnBoard Autonomy 2* (P*ROBA2*/SWAP), and NRL's Large Angle and Spectrometric Coronagraph on ESA's *Solar and Heliospheric Observatory* (*SOHO*/LASCO). For example, the southeastern CME is related to the solar flare whose origin we trace with a SWAP series of images.

## 1. INTRODUCTION

The properties of the white-light corona are governed by the large-scale magnetic field of the Sun, for which we have only these indirect measurements, though future missions including Solar Probe+ and Solar Orbiter are planned to sample closer to the Sun than previously possible. The shape of the white-light corona is sensitive to the phase of solar-activity cycle (Golub & Pasachoff, 2010; Pasachoff, 2009ab, Golub & Pasachoff, 2014).

We present here the results of the observations of the white-light corona (WLC) during the 2012 total solar eclipse. The eclipse appears in the catalogues as 13 November (UT), which translated to the early morning on 14 November locally for both observing sites reported here. Our main teams were stationed at the position of longest totality available from the ground (avoiding observations from the unstable platform of a ship), the northeast Queensland coast of Australia near Cairns and Port Douglas. We compare our results with images obtained from a ship (*Pacific Venus* in Figure 1) in the Pacific Ocean north of New Zealand to look for temporal changes. We also compare these results with the corresponding spaceborne observations. We reported preliminary results to the Solar Physics Division of the American Astronomical Society (Pasachoff et al., 2012).

Our comparisons are similar in method to those we reported from pairs of observing sites at the 2006 eclipse (Pasachoff, et al. 2007, 2008) from Greece; at the 2008 eclipse (Pasachoff, et al. 2009) from Siberia; at the 2009 eclipse (Pasachoff, et al., 2011b) from China; and at the 2010 eclipse (Pasachoff, et al., 2011a) from Easter Island. But, out of that series, only with this 2012 total solar eclipse did the Sun approach the maximum phase of the solar-activity cycle, so the corona was in a different configuration. As with the 2010 eclipse, our comparison data benefitted from an erupting Coronal Mass Ejection (CME). We again take advantage of observations made by the current generation of solar spacecraft, including instruments on NASA's *Solar Dynamics Observatory* (*SDO*), on JAXA's *Hinode*, and on ESA's *Project for Onboard Autonomy 2* (*PROBA2*) as well as full-sun coverage, including the whole far side, from NASA's *Solar Terrestrial Relations Observatory* (*STEREO*), which also supplied outer coronal views from its pair of perspectives.

## 2. BASIC INFORMATION ON THE 13/14 NOVEMBER 2012 ECLIPSE OBSERVATIONS

The path of totality started in the Australian outback and reached the northeast Queensland coast with the Sun at only 13° altitude. The rest of the path was entirely over the Pacific Ocean (Figure 1; Jubier, 2012ab; Espenak & Anderson, 2012; Zeiler, 2012; Golub & Pasachoff, 2014). No airborne expeditions intercepted the path except for our helicopter. Only passengers on ships saw totality after it left the Queensland coast and adjacent islands.

We chose our original observing site while relying on the cloud statistics from Anderson (2012, private communication) and Espenak and Anderson (2012, private communication). We noted also that the low altitude of totality, 13°, could lead to potential obscuration by clouds that does not show well on the satellite views that went into the statistical calculations.

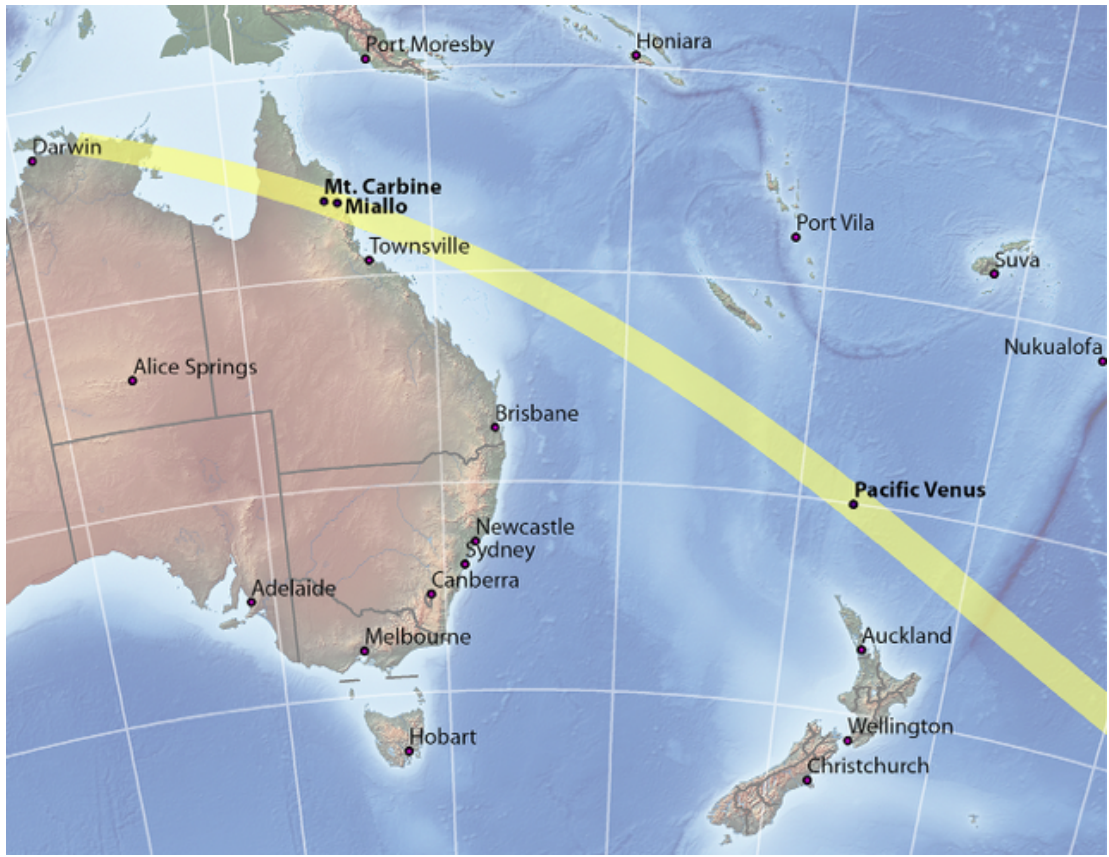

FIGURE 1. A map showing the path of totality of the 13/14 November 2012 eclipse (14 November, local date and time); ground-based observing sites such as Mt. Carbine and Miallo are located near the northeast coast in Queensland, Australia. (Courtesy: Michael Zeiler, eclipse-maps.com)

2.1. *Mt. Carbine*

Our observation site at Mt. Carbine was at 16°16′22″ S latitude and 144°42′53″ E longitude, on the Tablelands, inland from the Queensland coast, with 2 m 2 s of totality, centered at 20:38:49 UTC. Part of our observing team, headed by B.A.B., reconnoitered there the day before the eclipse and stayed the night there before the early morning eclipse, in view of the apparent cloudiness of our long-term Miallo site. The image shown in Figure 2 was made from fifty-eight individual frames from a RED Epic (www.red.com/products/epic) IMAX-quality high-resolution camera operated by R.D. and Nicholas Weber; we discuss the image processing by P.G., which included the use of darks and flats, in Section 3 of this paper. We have additional Nikon D90 + Nikkor 500-mm-lens frames taken by B.A.B.

Also at this last-minute site was a slitless spectrograph. Its data were recorded with a second RED Epic camera, to continue the studies of coronal temperature from the [Fe X]/[Fe XIV] ratio as previously reported by our group (Voulgaris et al., 2010, 2012). The RED Epic camera has a high dynamic range of 13.5 stops, up to 120 fps at 5k resolution with an effective 13.8-megapixel 14-bit CMOS sensor. The minimally compressed REDCODE RAW digital format preserves the original color space. We acquired the WLC and spectral images at 24 fps from C1 (1st contact) to C4 (4th contact). Relevant frames were adjusted for exposure and ISO values, then exported as TIFF files for image processing.

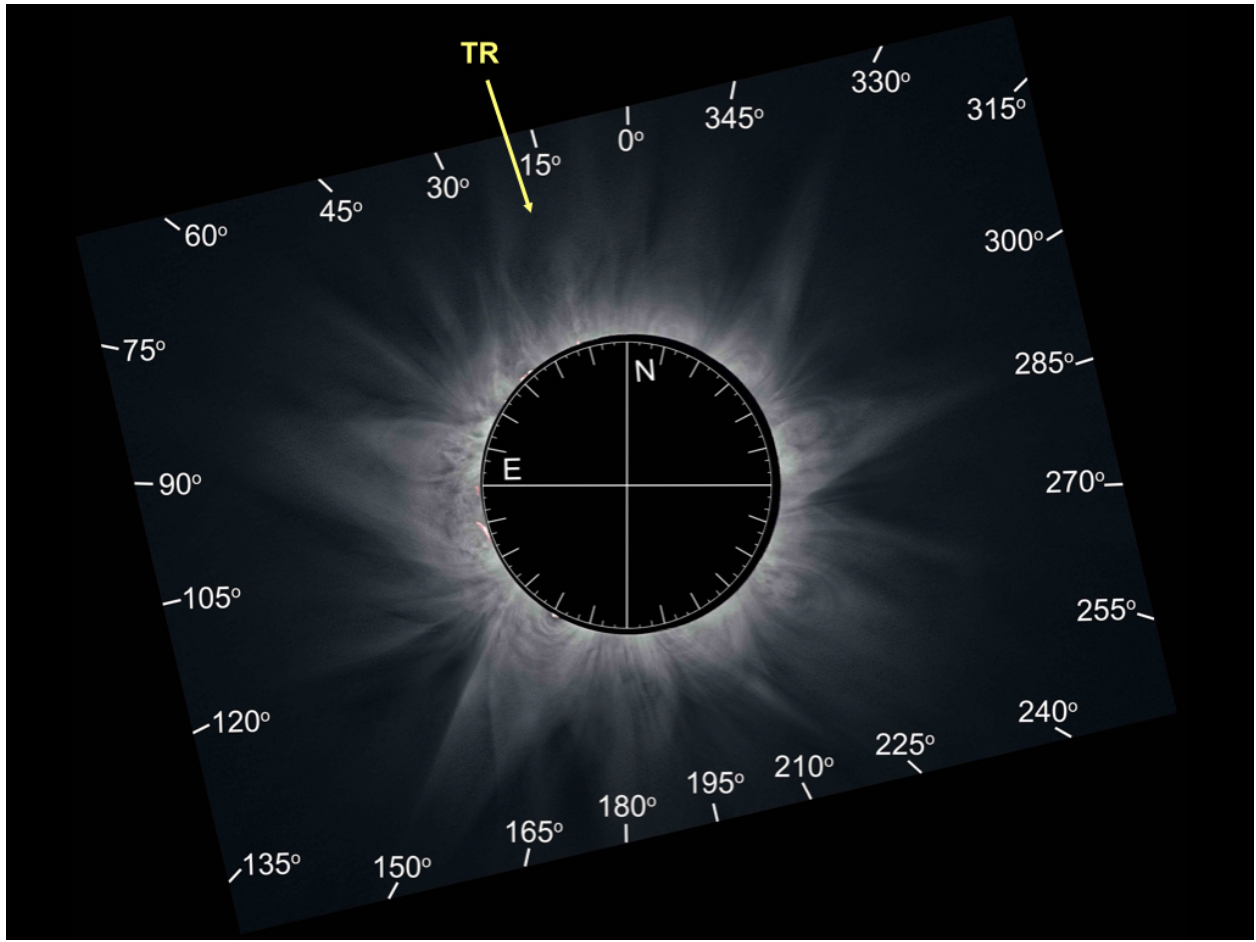

FIGURE 2. A combination of fifty eight white-light corona images, taken by R.D. and Nicholas Weber at Mt. Carbine, Queensland, Australia, computer-processed by P.G; solar north, east limb, and position angles (PA) around the solar limb are labeled in white.

2.2. *Miallo Observations*

On a reconnoitering trip a year in advance, we had found a house about 10 km inland from Newell in the town of Miallo. It was at a few hundred meters of altitude, to provide a sweeping view of the shadow's departure as well as the low-altitude totality above the ocean.

Observations at eclipse time in the early morning over the week preceding totality did not give hope for clear eclipse weather, so a group took much of our equipment inland, section 2.1 above. Some observers remained at Miallo in case of suitable holes in the clouds, which would allow the use of our best-calibrated equipment on the best-aligned mounts; that work was supervised by M.L.

When the weather looked definitively cloudy about a half hour before totality, two of us (J.M.P. and Robert Lucas joined by a BBC camera operator (a documentary about the Sun, *The Secret*

*Life of the Sun*, was in preparation; our work is in the international version but not the U.K. version) went aloft in a helicopter that we had arranged previously to stand by. We rose in circles above Miallo to a location above the 8000-foot cloud deck, reaching 9000 feet for totality (Figure 3). Images were obtained with a Nikon D3x and a Nikkor 80–400 mm VR zoom lens at its longest setting, the Vibration Reduction feature allowing a set of narrow-angle exposures that gave useful images.

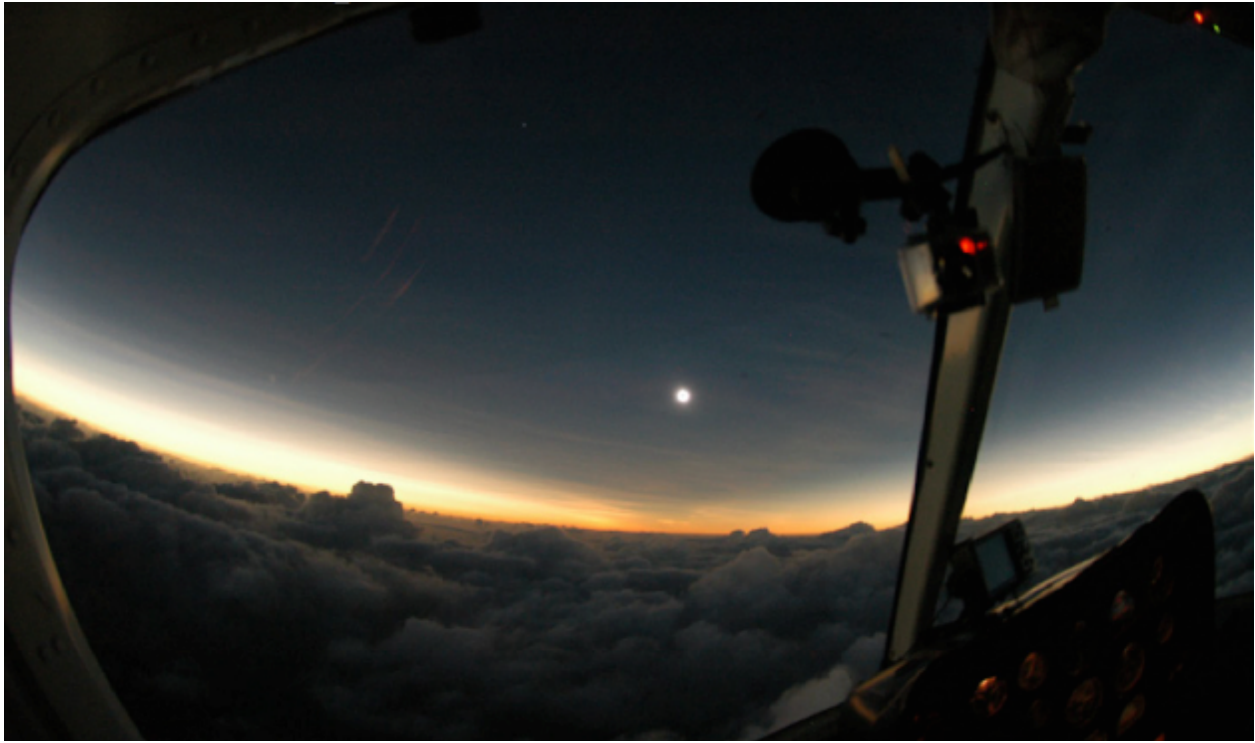

FIGURE 3. A wide-angle view from a helicopter over Miallo allowed the umbra to be seen racing across clouds; narrow-angle views were also obtained.

## 2.3 *Coastal Observations*

A group of 16 astronomers and astrophysics students from the Aristotle University of Thessaloniki, led by J.H.S. and A.V., with participation from University of Chicago scientist Thanasis Economou, were at an oceanside house at Newell, due east of Miallo. Equipment included a sophisticated spectrograph (constructed by A.V.) for the temperature measurements (Voulgaris et al., 2010, 2012). A satellite site was 20 km north along the coast. Clouds prevented observations from these sites.

Several colleagues had equipment at Trinity Beach, just north of Cairns. Some images by people in our group were obtained through holes in the clouds (Amy Steele, Michael Kentrianakis, Aram Friedman), as well as videos (Friedman, 2012) that showed the duration of the clear intervals.

Other professional teams, including that of Sterling, of Druckmüller, of Habbal, and of Koutchmy, were in adjacent apartments.

Ten km north along the coast, a meeting of astronomers was held at Palm Cove. A hole in the clouds drifted over their location for totality, and most were fortunate to see the corona then.

2.4 *Shipborne Observations*

The main set of images with which we compare our Queensland data were obtained by Kazuo Shiota from the *Pacific Venus*, the second largest cruise ship (~800 passengers) registered in Japan. He observed at 21:15 UTC from E 173°4.4′ and S 30°2.2′. He obtained 36 good-quality images in which the corona was fully shown, limited by the ship's rolling from the 100 images he took in total. He also took darks and flats.

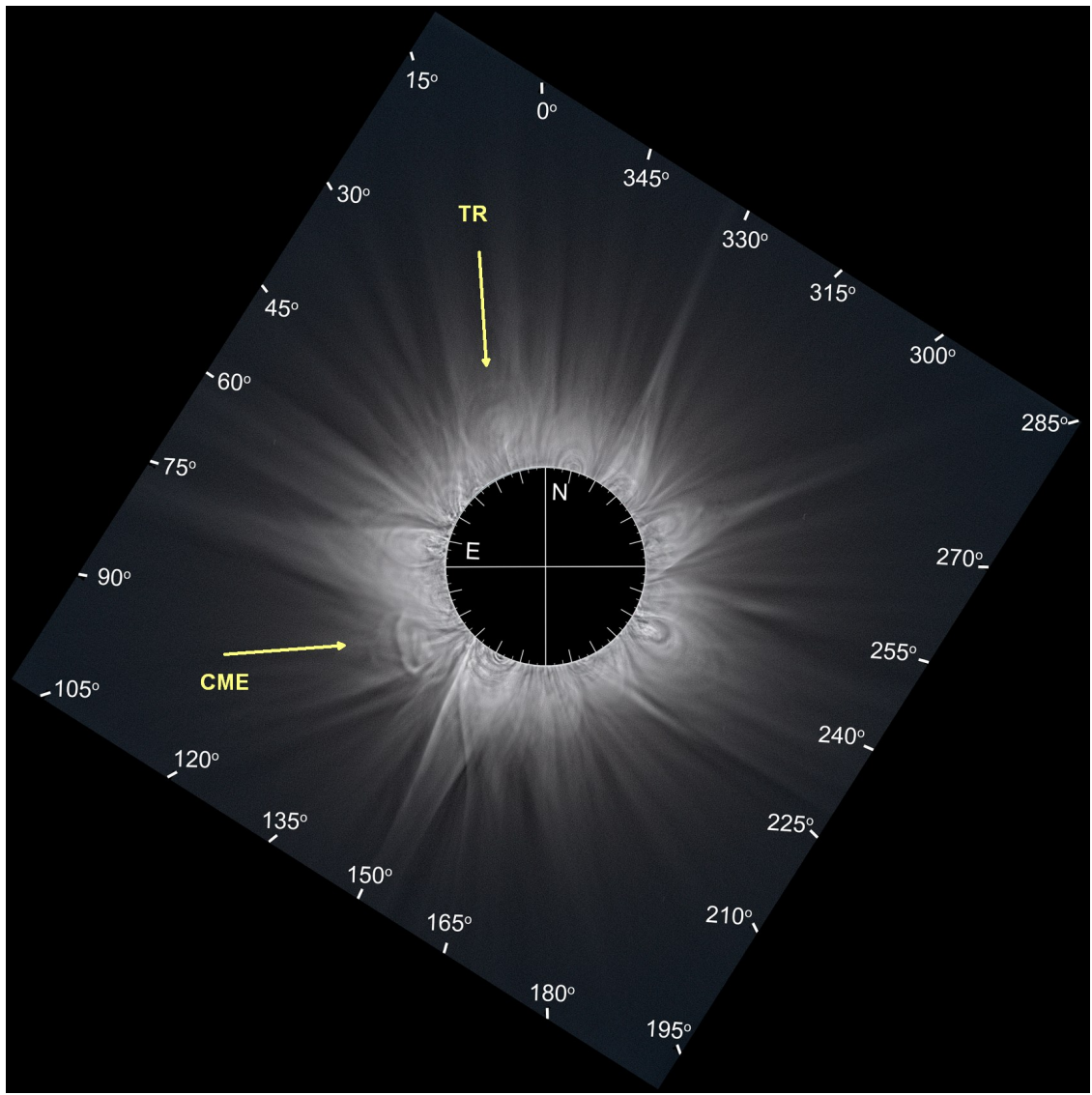

FIGURE 4. A combination of 36 white-light corona images, taken by K. Shiota from a ship north of New Zealand, computer-processed by P.G.; solar north, east limb, and position angles (PA) around the solar limb are labeled in white color; a CME, not observed in Figure 2, is marked with yellow arrow; a tennis-racquet shaped structure is labeled TR. A GIF blinking unlabeled versions

of Figures 2 and 4 is available in the Supplemental Material. Hanaoka et al. (2014) have also analyzed a pair of images from this eclipse; the pair they studied is separated by 35 minutes.

## 2.5 *Methods of Image Processing*

Fifty-eight images from the RED Epic camera on the Tablelands were combined into a composite image that shows the solar-maximum structure of the solar corona shown in Figure 2. Though manual, our reduction method was similar to that used by Druckmüller, Rušin, & Minarovjech (2006), Druckmüller (2009), Druckmüllerová, Morgan, and Habbal (2011), and Druckmüller, Habbal, and Morgan (2014); Druckmüller's system was unable to handle the REDCODE RAW format of our RED Epic images.

Our basic processing steps were:
1) The coordinates of the center of the Moon were detected with high accuracy. The process is automatic and it is based on mathematical morphology (Petrou & Sevilla, 2006).
2) A radially graded filter from the center of the Moon was applied to the fifty-eight images. This computer filter removes the steep decrease of brightness of the solar corona and keeps the corona's structure unchanged.
3) The images of step 2 were aligned, using a phase-correlation method that is based on 2D Fourier transforms. The accuracy of alignment of the above method is ±0.5 pixels. After this step the alignment translations ($\Delta x$, $\Delta y$) were calculated for each image.
4) The information of step 3 was used in order to align the pure (unprocessed) RAW images.
5) The aligned images were combined. The weights were proportional to the exposure time of each image.
6) Finally, in order to enhance the combined image, locally adaptive filters were applied.

## 3. PROMINENT FEATURES OF THE STRUCTURE OF THE WHITE-LIGHT CORONA

The brightness of the white-light solar corona is a result of Thomson scattering of the photospheric light on free electrons. However, distribution of free electrons is maintained by solar magnetic fieldlines, both local and global, which extend from the photosphere to the solar corona and create different coronal structures. Helmet streamers are the most important for the brightness distribution; during solar maximum they are nearly uniformly distributed around the solar limb. This type of 'corona maxima' has flattening index below 0.1 and is very rare. The biggest difference in coronal brightness between solar maximum and minimum is around cycle minima, because helmet streamers are then localized only around the solar equator. For such 'corona minima' the flattening index is largest (around 0.3). The flattening index is a very simple coronal parameter to show the overall distribution of magnetic fields with latitude on the Sun. We note that classical helmet streamers are localized above the neutral lines that separate opposite polarities of solar large-scale magnetic fields. The distribution of helmet streamers and their dynamics over the solar-activity cycle was shown by Seagraves et al. (1983) and Belik et al. (2004).

As a first approximation, the 13/14 November 2012 white-light corona seen on our combined image can be regarded as of the solar-maximum type with a number of helmet streamers located around the whole solar disk, of varying base-width and different orientation with increasing height above the surface (Figures 2 and 4).

Habbal et al. (2010a), Habbal et al. (2011abc), Habbal et al. (2012), and Habbal et al. (2013) have also used similarly processed images, from the 2006, 2008, and 2010 eclipses, as part of their investigation of emission-line ratios and the transition from collisional to collisionless plasma. Habbal et al. (2010b) have reported about such coronal fine structure as prominence shrouds. Habbal et al. (2013) and Habbal et al. (2011b) have discussed methods of probing coronal physics with total eclipse observations.

Our images provide the latest value of the flattening index, $\varepsilon$, combining the radial distances as shown in Figure 6a and called $E_i$ (for Equatorial structure) and $P_i$ (for Polar structure):

$$\varepsilon = \frac{E_1 + E_2 + E_3}{P_1 + P_2 + P_3} - 1,$$

(Özkan, M. T. et al., 2007, in their discussion of the 2006 eclipse coronal flattening), which measures the flattening index of coronal isophotes at 2 $R_\odot$. Our measured value of 0.01 (Figure 6b) fits the curve for this phase of roughly 0.07 of the solar-activity cycle (Golub & Pasachoff, 2010, and references therein, Figure 6); the graph of the flattening index, shown there (Figure 4.11) was updated based on information from S. Koutchmy, V. Rušin, & M. Druckmüller (private communications). It is obvious from Figure 6b that the flattening index of the isophotes shown in Figure 6a decreases as we move away from the solar limb. This effect reflects the changing structure of the magnetic field of the Sun as a function of distance away from the photosphere. The flattening index we measure fits well with the phase of the solar-activity cycle (Figure 7).

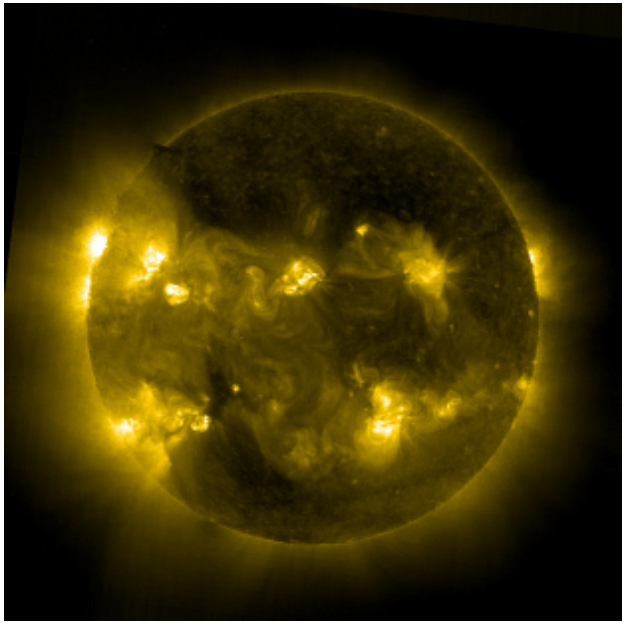

FIGURE 5. An image of the 284 Å EUV corona (Fe XV, 2 × 10⁶ K), reprocessed by D.B.S. This EUV corona is almost absent in the northern hemiphere at position angles (PA) 295°–45°, even though the white-light corona structures are there well discernible. (Courtesy: ESA/NASA/*SOHO*/EIT)

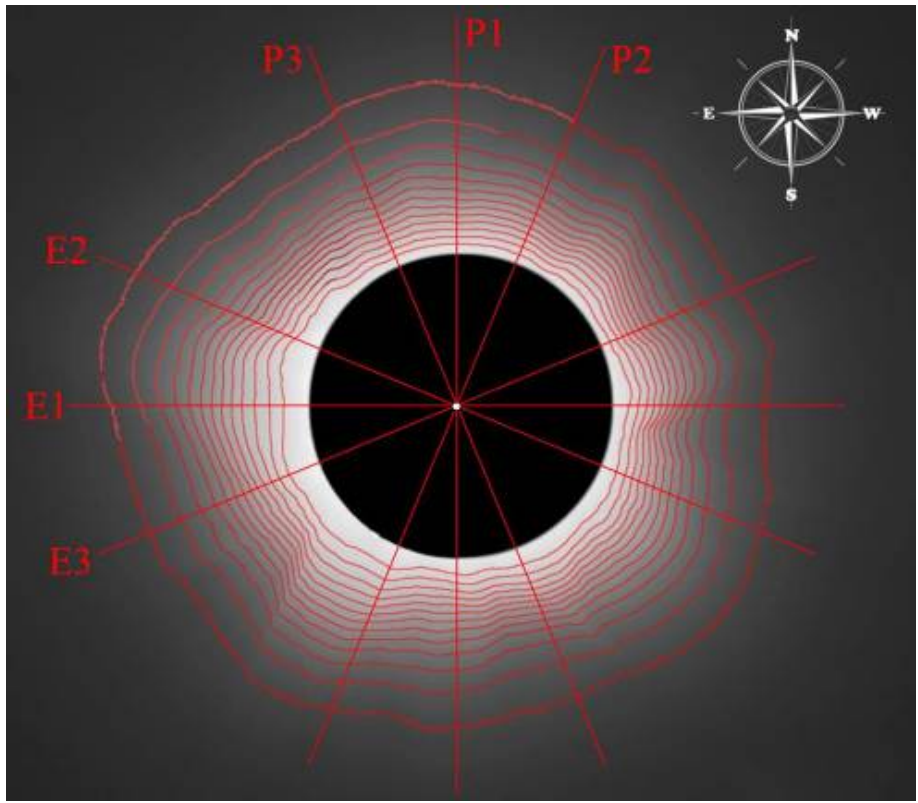

FIGURE 6a. Isophotes superimposed on our composite image.

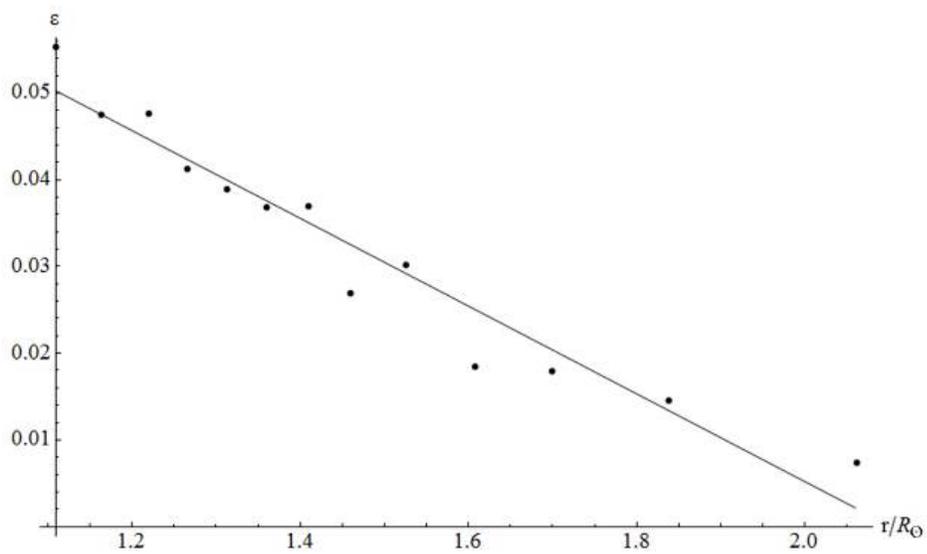

FIGURE 6b. The flattening index of isophotes as a function of solar radius.

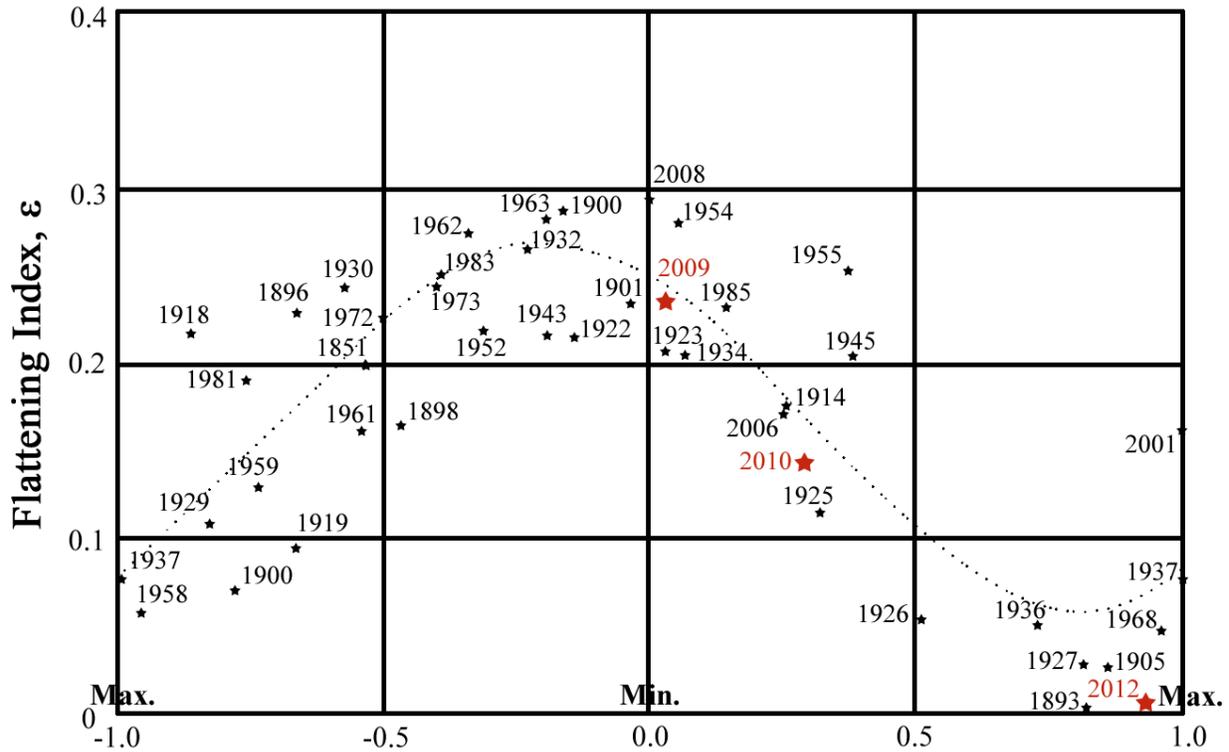

FIGURE 7. Variation of the flattening index over the phase of the sunspot cycles, with the large, red asterisks singling out the 2009 and 2010 observations made by the authors' recent publications in this journal and the 2012 point from the current paper. Even when the magnetic field strength at the solar poles, as measured by Wilcox observatory (http://www.solen.info/solar/polarfields/polar.html), is nearly twice lower in comparisons with previous solar cycles since 1976, the flattening index around cycle maxima has nearly the same value. (Updated from Golub and Pasachoff 2010; see further analysis of the evolution of coronal flattening in Pishkalo 2011.)

## 4. DYNAMICS OF THE WHITE-LIGHT CORONA AT SOLAR MAXIMUM

Short-term changes of small-scale structures of the solar corona have been lately, thanks to spaceborne observations, quite intensively studied, e.g., Sheeley et al. (2007), Sheeley and Wang (2007), Moreno-Insertis, Galsgaard, & Ugarte-Urra (2008), and van Ballegooijen, et al. (2014). The behavior of large-scale solar coronal structures have been discussed, for example, by Rušin and Rybanský (1984), Koutchmy (1988), Zirker et al. (1992), Pasachoff et al. (2006, 2007, 2008, 2009, 2011ab), Golub & Pasachoff (2010, 2014), Habbal et al. (2013), and Druckmüller et al. (2014). As the observing sites (Queensland and north of New Zealand) of the images that are going into our comparison were 36 minutes apart in umbral travel time, comparing the corresponding data also enables us to discern interesting changes in the large-scale structure of the white-light corona on a temporal scale of a half hour. We shall briefly comment on four cases.

The 2012 eclipse fell on a high plateau between the two peaks of cycle 24 (Figure 8): late 2011/early 2012 and ongoing during 2014, according to the mean monthly sunspot number (SILSO, World Data Center – Sunspot Number and Long-term Solar Observations, Royal

Observatory of Belgium, on-line Sunspot Number catalogue: http://www.sidc.be/SILSO/, "1950-2014"). The level of activity of the Sun was relatively high, compared to that of the eclipses toward the end of the 23$^{rd}$ cycle, though lower than at other eclipses during sunspot maxima of the past century. Figure 9, which contains both a magnetogram image from HMI and a 171 Å image from AIA (both on *SDO*), shows the magnetic-field solar conditions during the eclipse.

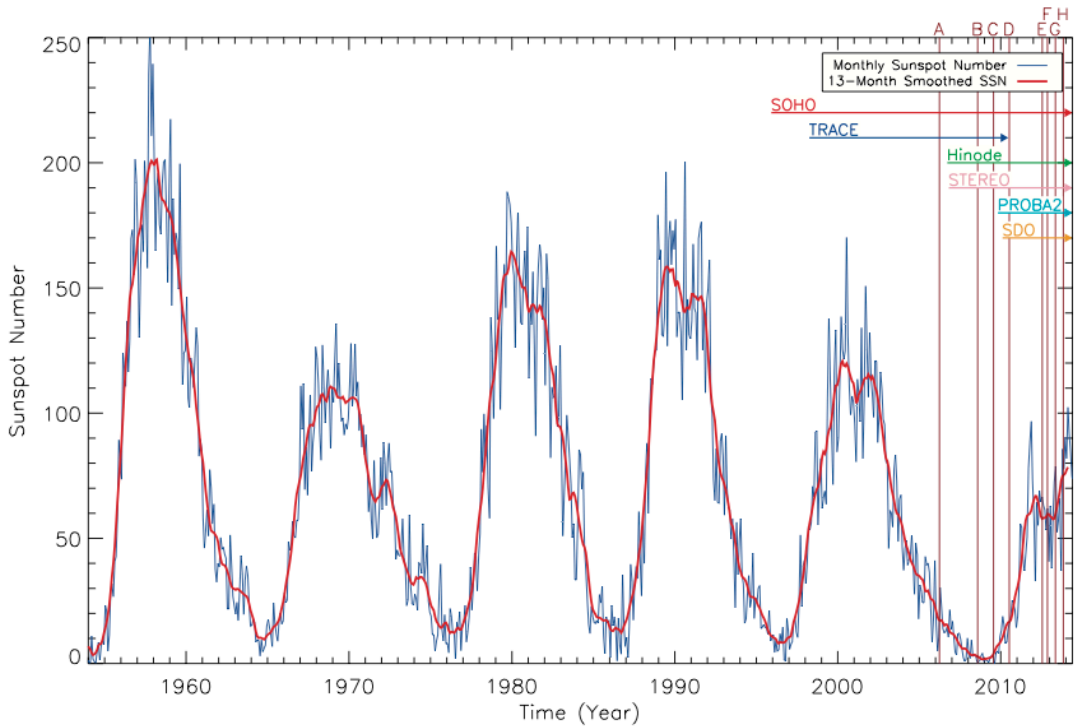

FIGURE 8. The sunspot cycle over the last several decades (cycles 19-24). The lifetimes of selected solar space observatories are indicated with vertical dotted lines. Solar eclipse dates are indicated with letter A to H. TSE stands for Total Solar Eclipse and ASE stands for Annular Solar Eclipse. A: TSE 2006 Mar. 29, B: TSE 2008 Aug. 1, C: TSE 2009 Jul. 22, D: TSE 2010 Jul. 11, E: ASE 2012 May 20, F: TSE 2012 Nov. 13, G: ASE 2013 May 10, H: TSE 2013 Nov. 3.

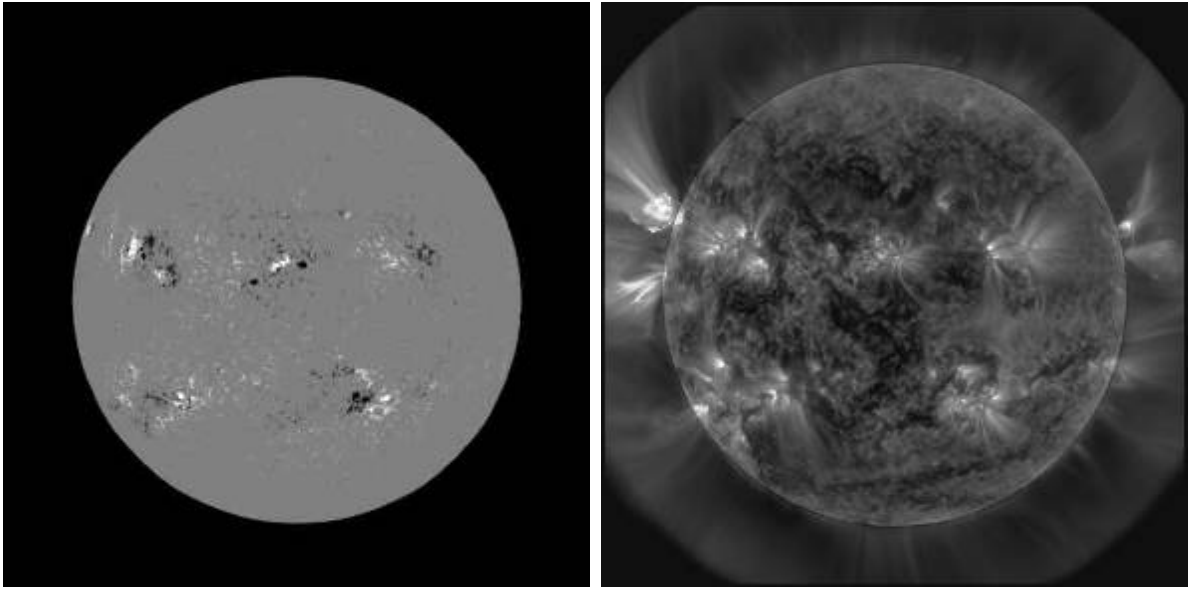

FIGURE 9. (a) Left: An *SDO*/HMI magnetogram corresponding to eclipse time in Australia. Terrestrial north is up. (b) Right: An *SDO*/AIA image at 171 Å showing the high-temperature Fe IX/X corona on the disk corresponding to eclipse time in Australia. The image has been enhanced with a radial filter. Terrestrial north is up.

The coronal compound images from Queensland and the NZ ship can be understood in terms of the underlying magnetic field. In Figure 10, we see a computed plot of the coronal magnetic-field lines, with source surface (where the magnetic field is constrained to become radial) at heliocentric distance 2.5 $R_\odot$, as Wang, Sheeley, & Rich (2007) described for their similar work at an earlier eclipse (see also Schatten et al., 1969). The extrapolation method is explained in Wang and Sheeley (1992). All field lines that cross the source surface are defined to be "open," with their footpoint areas representing coronal holes. Their calculation used NSO/Kitt Peak photospheric field maps for Carrington Rotation (CR) 2129 and CR 2130 (October – November 2012); Mt. Wilson Observatory and Wilcox Solar Observatory photospheric maps gave very similar results (Figure 11). Such Carrington synoptic maps are assembled from central meridian observations taken over the given Carrington Rotation (Figure 12), and reproduced here for long-term reference. The map thus includes data taken both before and after eclipse day. This hairy-ball plot gives a general idea of the topology of the coronal field on November 13/14, showing the locations of coronal holes, helmet streamers separating holes of opposite polarity, and "pseudostreamers" (Wang, Sheeley, & Rich 2007) separating holes of the same polarity. For a review of calculations of the sun's global magnetic field, see Mackay (2012). Habbal et al. (2011b) have further discussed linking eclipse observations with determining the structure of the coronal magnetic field while Judge, Habbal, and Landi (2013) discuss line-of-sight and other difficulties and properly interpret coronal observations. Alternative calculations by Mikić (2012 – "Predictive Science" http://www.predsci.com/corona/nov2012eclipse/nov2012eclipse.html) predicting the coronal appearance were posted prior to the eclipse (Figure 13a), allowing verification of the validity of his calculational system according to that usual scientific-method test. The predicted model (Figure 13b) was very close to the real state of the white light corona (WLC). Therefore, such method should be very useful for modeling of the solar corona and forecasting of solar wind distribution over a solar cycle. Additionally, the high resolution WLC images, obtained from ground observations during solar eclipses, remain the best tracers of the global coronal magnetic field structure.

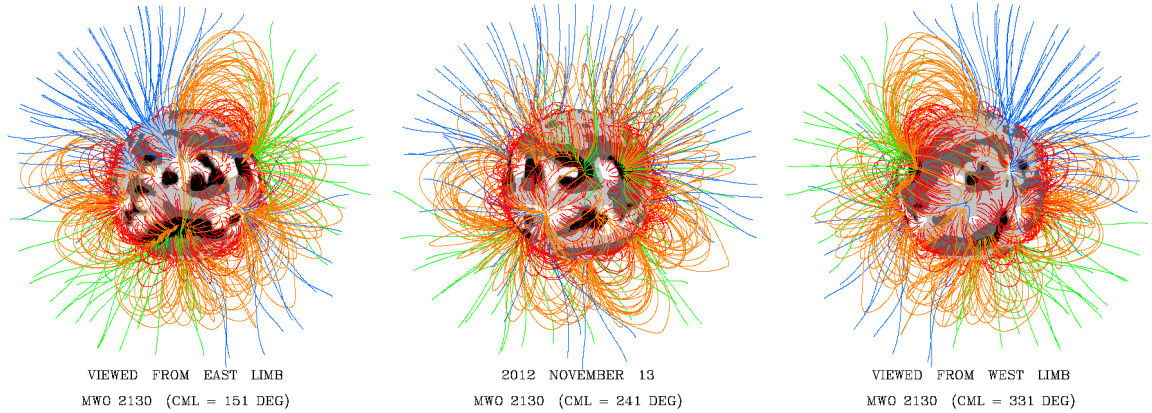

FIGURE 10. A completed plot of the coronal magnetic-field lines, calculated for a source surface at heliocentric distance 2.5 $R_\odot$ by Y.-M. Yang (NRL), and viewed from the east, from the Earth, and from the west. Open field lines are depicted in blue (outward-directed) and green (inward-directed) field lines, with closed field lines in orange (long loops) and red (short loops).

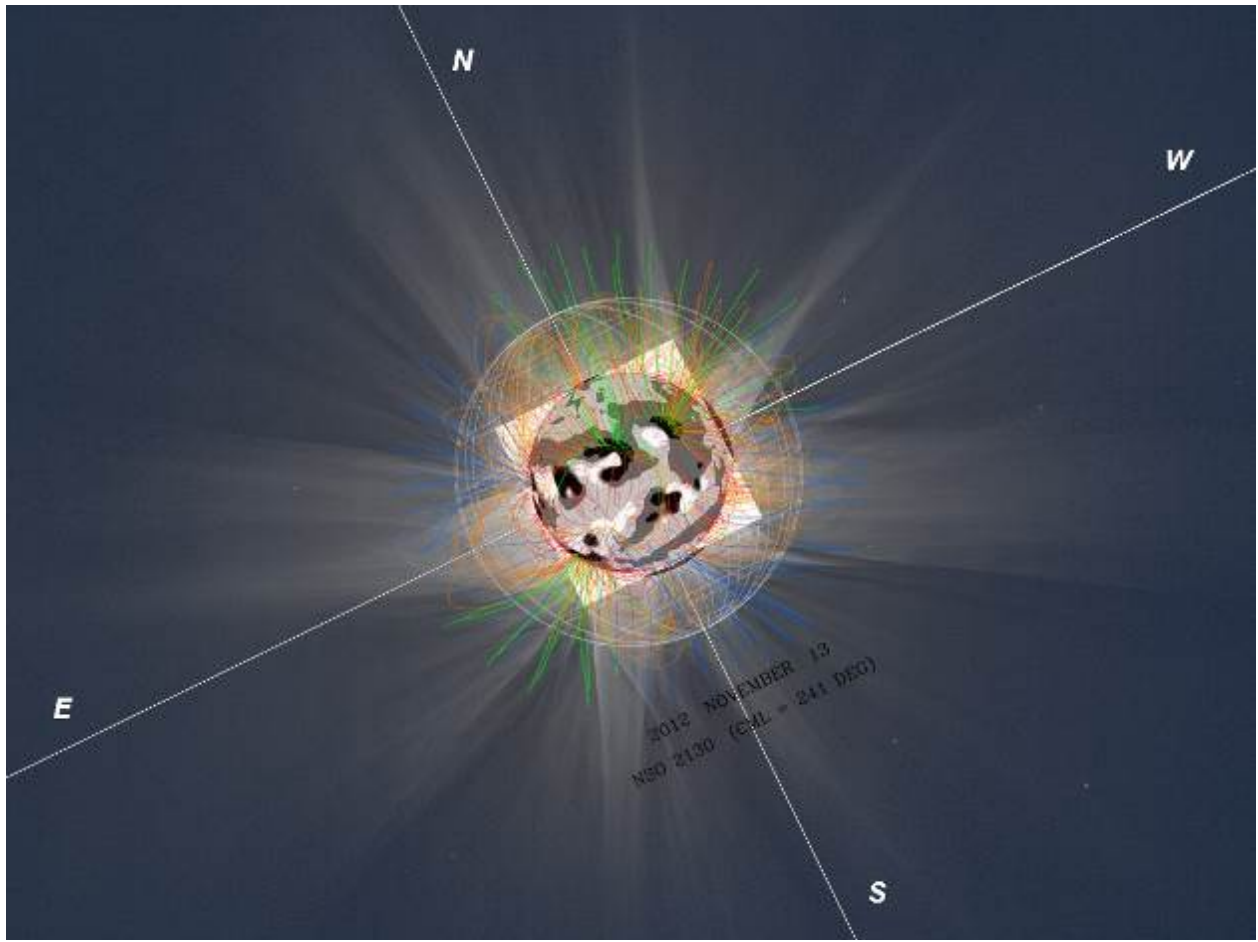

FIGURE 11. The coronal magnetic-field calculations of the previous figure overlain on a composite eclipse image by M. Druckmüller from images by several contributors (http://www.zam.fme.vutbr.cz/~druck/eclipse/Ecl2012a/0-info.htm).

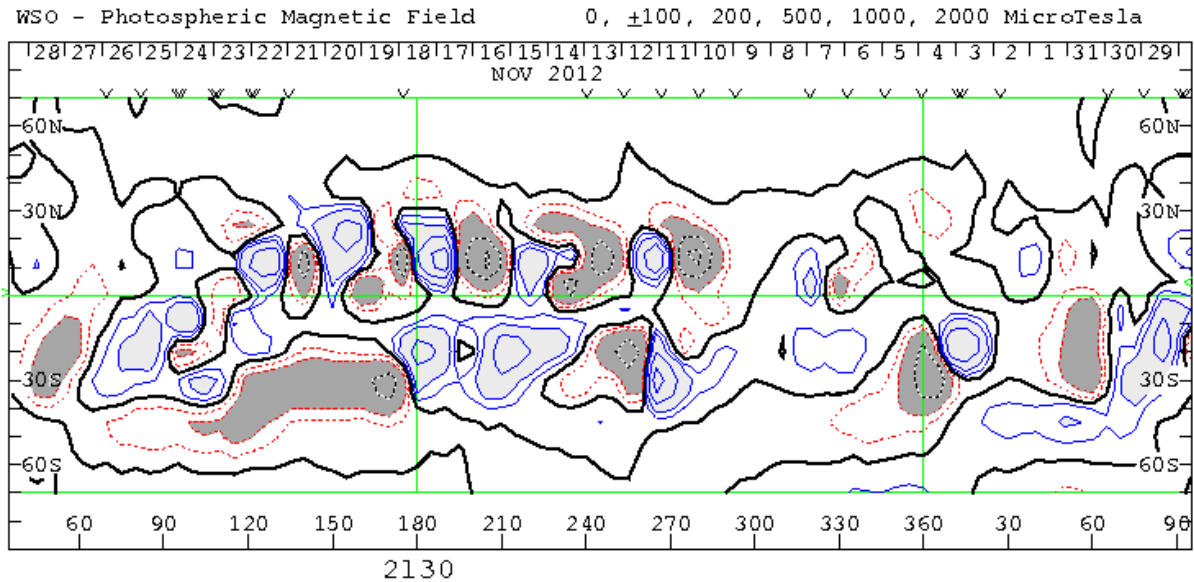

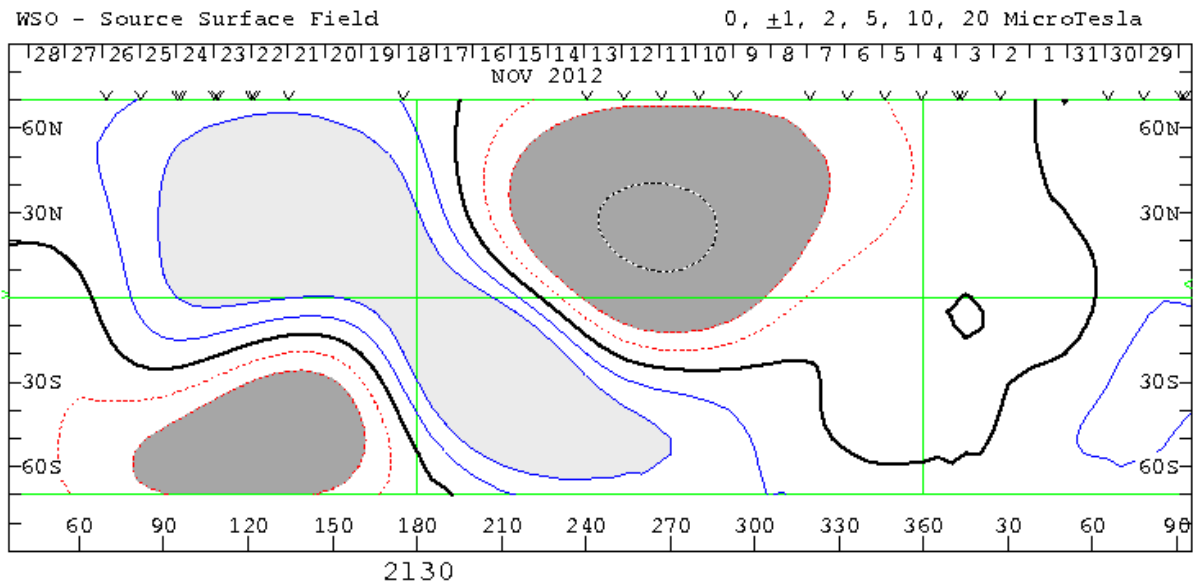

FIGURE 12. (a) Top: Synoptic charts of the solar magnetic field are assembled from individual magnetograms observed for the CR 2129–2130 graphed for the month centered on the eclipse day. The contour maps show the distribution of magnetic flux over the photosphere. Blue, light shading shows the positive regions (outward field lies). Red, dark shading shows the negative regions (inward field lines). The neutral line is black. (b) Bottom: The computed Wilcox Observatory coronal magnetic-field map (the source surface at 2.5 solar radii) for the Carrington Rotation 2129–2130 graphed for the month centered on the eclipse day. Blue, light shading shows the positive regions (outward field lies). The neutral line is black. Red, dark shading shows the negative regions (inward field lines). (Courtesy of Todd Hoeksema, Stanford U.)

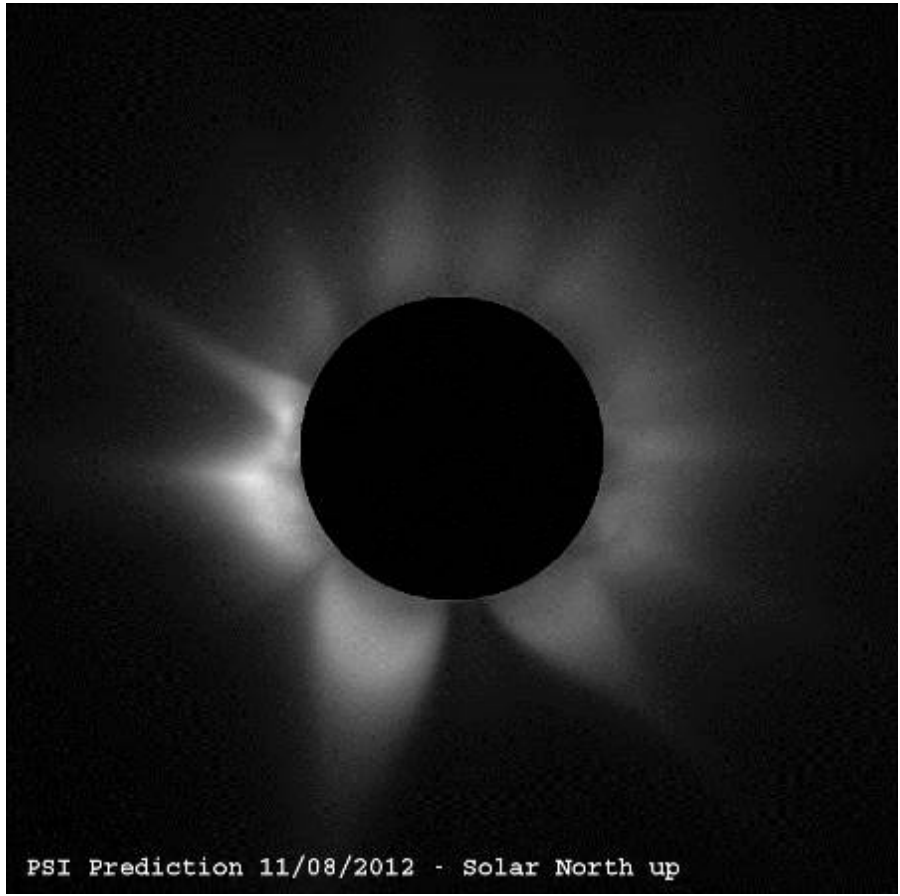

FIGURE 13. (a) The pre-eclipse prediction of the pB (polarized brightness) coronal structure based on *SDO*/HMI data, posted five days before the eclipse by Z. Mikić (Predictive Science).

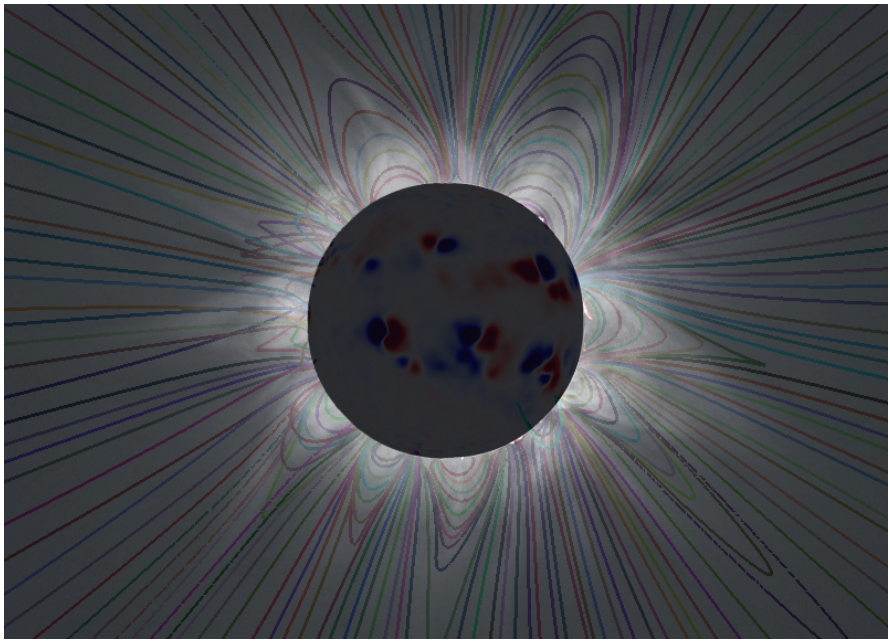

FIGURE 13. (b) Visual comparison between our group's composite image from the eclipse observation (background) and model-predicted coronal magnetic field (overlay with 15% opacity) shows that the used model is relatively good; however, computation of magnetic coronal structures from solar surface observations of magnetic field is very complicated and more models (computations) should be used. Solar north is 169º counterclockwise from the vertical. Magnetic-field caculations and overlay by Zoran Mikić and Jon Linker, Predictive Science, Inc.

We also compared our images of the white-light corona with EUV images obtained using the *Sun Watcher with Active Pixels and Image Processing (SWAP)* onboard ESA's *PROBA2* spacecraft (Seaton et al., 2011; Seaton et al., 2013, Halain et al., 2013). SWAP images have a passband with its peak at 174 Å and containing the Fe IX/X emission lines that form near 1 million K. In order to improve the signal-to-noise ratio of the SWAP images at large distances above the solar surface where the EUV corona is very faint we generated two composites of fifty 10-s images that were obtained during two 60-minute windows surrounding the pair of ground-based eclipse observations. The SWAP composite corresponding to the eclipse observation is shown in Figure 14. In order to reduce the large dynamic range of the composite, the part of the image outside the limb has been treated with a radial filter to remove some of the overall falloff in intensity from the bright inner corona to more extended structures.

While the *SDO*/AIA 171 Å passband includes mainly Fe IX with a large contribution from Fe X (Lemen et al., 2011), the *PROBA2*/SWAP 174 Å passband includes mainly Fe X with a large contribution from Fe IX. Because SWAP's passband is wider than AIA's, it is useful for detecting the flare emission lines of Fe XX and Fe XXIV lines (AIA's passband is therefore dominated by the ~ MK gas shown by Fe IX/X). The differences among various instrument responses near 171 Å for *SOHO*/EIT, *STEREO*/EUVI, *SDO*/AIA, and *PROBA2*/SWAP are discussed in Raftery et al. (2013). Habbal et al. (2012) compared the use of such EUV lines with visible-spectrum forbidden emission lines, finding some advantages for the latter, when calibrated, as diagnostic tools, with further mention of possible future use of the visible forbidden lines by Judge et al. (2013). We will discuss our forbidden-line spectra from the 2012 and 2013 total solar eclipses in subsequent papers, continuing the series of Voulgaris et al. (2012).

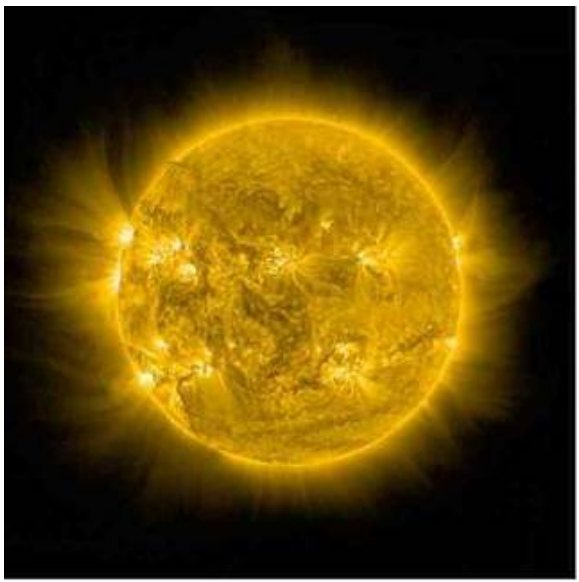

FIGURE 14. (a) A composite of fifty 10-s SWAP images acquired during a 60-minute window surrounding the eclipse observation, showing the full extent of the corona as seen in SWAP's passband, which is centered at 174 Å (0.8–1.0 MK).

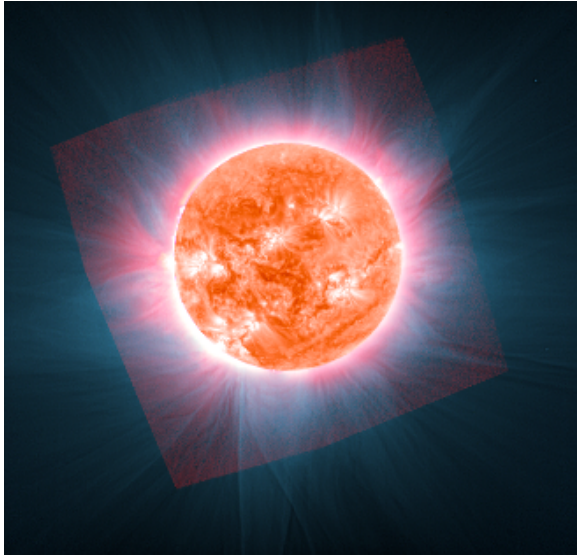

FIGURE 14. (b) *PROBA2*/SWAP composite image of the corona (red), superimposed on white-light corona image of the Queensland eclipse (blue), showing the on-disk sources of several coronal features. (*PROBA2*/SWAP Consortium/Royal Observatory Belgium)

A sample of AIA images appears in Figure 15; a full set is in the Supplemental Material.

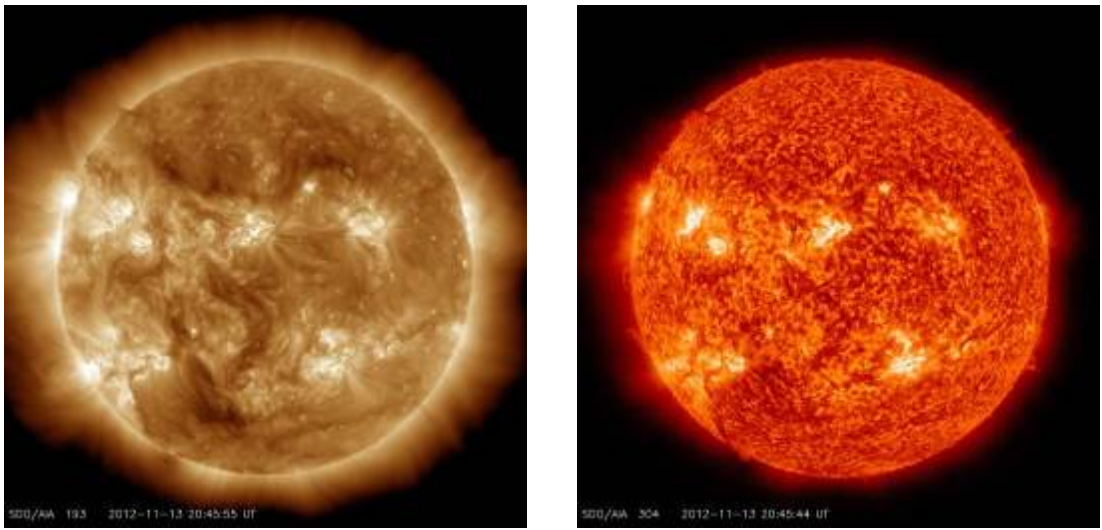

FIGURE 15. (a) Left: Spatial distributions of active regions seen in Fe XII at 193 Å, 1,600,000 K and (b) Right: He II at 304 Å (right) observed with NASA's *SDO*/AIA at the time of the eclipse. Compare with the HMI image of the magnetic field in Figure 9a. A full set of *SDO*/AIA images from close to eclipse time is available in the Supplemental Material. (Courtesy: *SDO*/LMSAL/NASA)

## 5. COMPARISON OF THE WHITE-LIGHT CORONA WITH SPACEBORNE OBSERVATIONS – A DETIALED TRACING OF A CME

The white-light corona of the total eclipse of 13/14 November 2012, can aptly and succinctly be described as a highly dynamic and intricately structured corona of maximum type (Figure 16). As mentioned above, the flattening index, ε, is only 0.01, one of the lowest values on a long-term scale (1851–2010) as studied by Pishkalo (2011). A preliminary estimate of the phase of the 24th cycle (last minimum: 2009, next minimum based on 11-year cycle: 2020) is then 0.37.

A number of helmet streamers are visible, some also well discernible from LASCO C2 and C3 (Brueckner et al., 1995) and evenly distributed around the solar disk. Their bases exhibit a great variety of bright and dark loops and arches, some of them extending far away from the solar limb, like those located at position angle (PA) around 80° and close to 260°. The most pronounced and extended system of helmet streamers is located at PAs ranging from 143° to 177°, overlaying a quiescent prominence and a coronal cavity seen in the 284 Å corona (see Figure 5) as well. The increased dynamics of the corona is illustrated by a number of truly spectacular classical CMEs in the period around the time of eclipse. Observations from C2 and C3 reveal a CME on 12 November at 00:12 UTC and at PA 300°, a CME on 13 November in the morning at PAs 119°–150°, and a couple of CMEs on 14 November, one at 2:24 UTC at PA 250° and the other at 12:48 UTC at PA 90°. SWAP images show the development of one of the CMEs on the visible disk in the EUV (Figure 17a and 17b). Morgan and Habbal (2010b) have discussed distinguishing CMEs from the quiescent corona. Koutchmy et al. (2008) have followed a limb CME outside of an eclipse.

The dynamical features of the white-light corona can be discerned by comparing our eclipse observations from Australia at 20:38 UTC and shipborne observations close to New Zealand (S 30° 02.0′, E 173° 04.4′) at 21:15 UTC (Figures 18a and 18b, processed from Figures 2 and 4 respectively). Whereas our observations from Australia show no CME, those made near New Zealand 36 minutes later do show, although a rather weak, CME in the interval of PAs ranging from 132° to 141°. This CME is also of a rather complicated shape. Its forerunner, located at about 1.89 $R_\odot$ (680,000 km) above the solar limb, is followed by three "legs" emanating from the solar limb at PAs 132°, 136°, and 141°. On the original images, this CME is already seen at 1.79 $R_\odot$ (550,000 km); from C3 this object can still be traced on 14 November at 4:30 UTC at as far as 23 $R_\odot$, moving at a speed of 200 km s$^{-1}$. This particular CME was probably related to a weak flare observed at 21:00:03 with SWAP (Figure 17a and 17b); the speed of the ejected mass was about 413 km s$^{-1}$ (Figure 17c). In the past, such CMEs have probably been mistaken for eclipse comets (e.g., Ranyard, 1879; Cliver, 1989).

It is worth mentioning that in the period around the eclipse time this particular region exhibited several mass ejections—this one is #65 in ROB's CACTus (Computer Aided CME Tracking); other CMEs associated with this eclipse are #58-70 (see http://sidc.oma.be/cactus/catalog/LASCO/2_5_0/2012/11/latestCMEs.html)—whether in a form of CMEs or propagation of brightening along plumes/arches. Often, as here, CACTus underestimates velocity, as we measured. The median velocity is decreased by a cluster of slower-moving points at larger position angles, but the core of the CME (closer to 90°) is moving faster, consistent with our measurements in the lower corona in the EUV.

Another notable object is of a shape reminding a table-tennis racquet (TR). Its "handle," looking like a screw, projected from a faint prominence at PA 19°. The bottom part of the arc is located at about 250,000 km and the upper part of the arc at 500,000 km above the solar limb. The highest point of the other loop, located above the described feature, is at 620,000 km, but its legs are rooted at different position angles. A comparison of the images taken at 20:38 UTC and 21:14 UTC shows no remarkable changes in its structure. However, observations from C2 (Brueckner et al., 1995) do not show any notable dynamics in this region up to midnight of November 13, 2012. It is also worth noticing a rather intricate complex of features around PA 242°. Well-developed dark and bright loops are discernible up to 2 solar radii, even though a typical helmet streamer is missing. Moreover, these loops seem to be overlaid by faint streamers, easily seen in the *SOHO* corona (see Figure 19 and 20). The orientation of the streamers with respect to the prominences can be seen at Emmaoulidis and Druckmüller (2012).

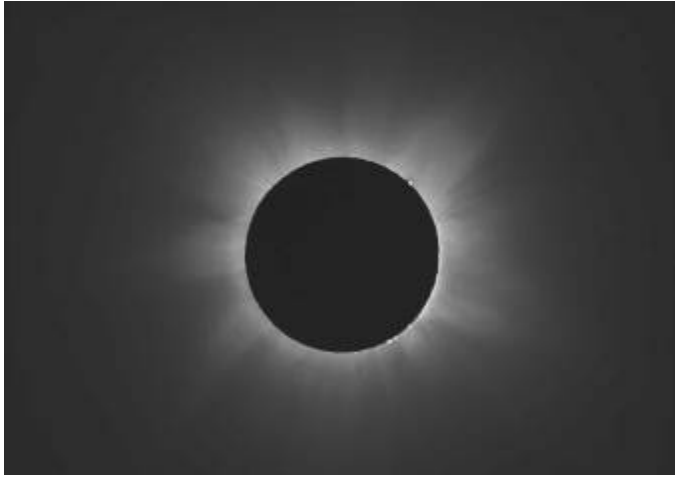

FIGURE 16. A composite of dozens of our individual images from the Mt. Carbine, Queensland, made by Wendy Carlos with less emphasis on contrast than the earlier composites (Figures 2 and 4) though clearly showing the low-flattening of this solar-maximum corona.

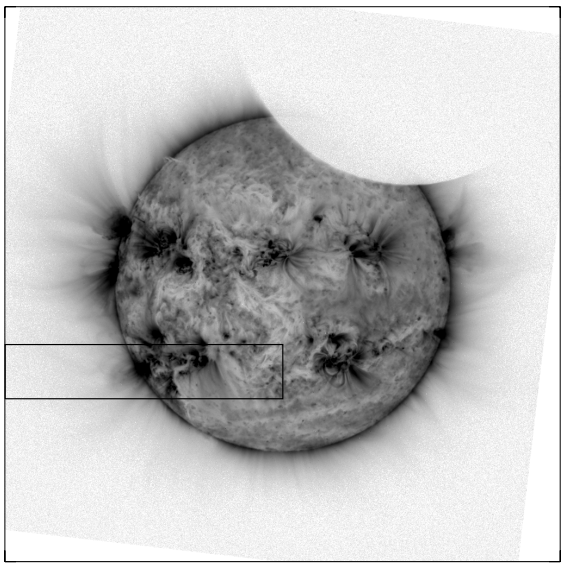

FIGURE 17a. A SWAP image taken at UTC 21:02 showing the 174-Å corona; inverted grayscale is used enhance contrast; the narrow box at the SW limb indicates the area sampled for a CME dynamics study (see Figure 17b and c).

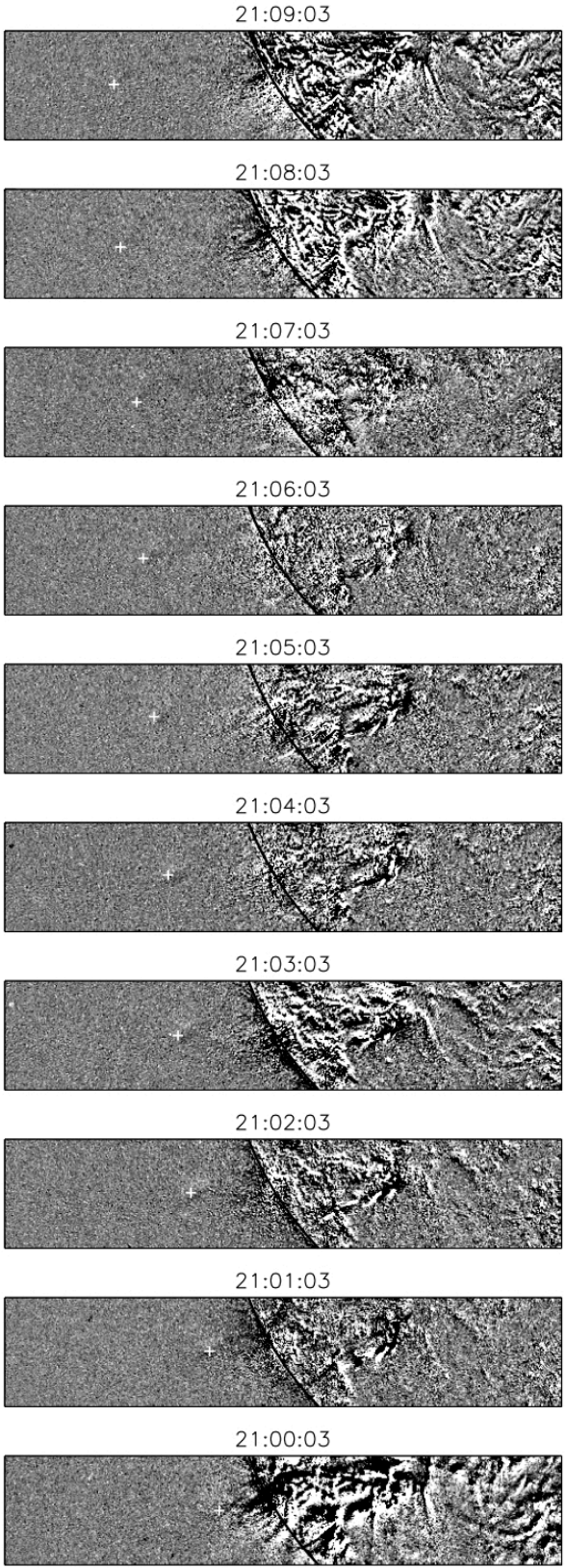

FIGURE 17b. A series of SWAP running-difference images showing the evolution of the CME on the disk; the area sampled is enclosed with a box in Figure 17a.

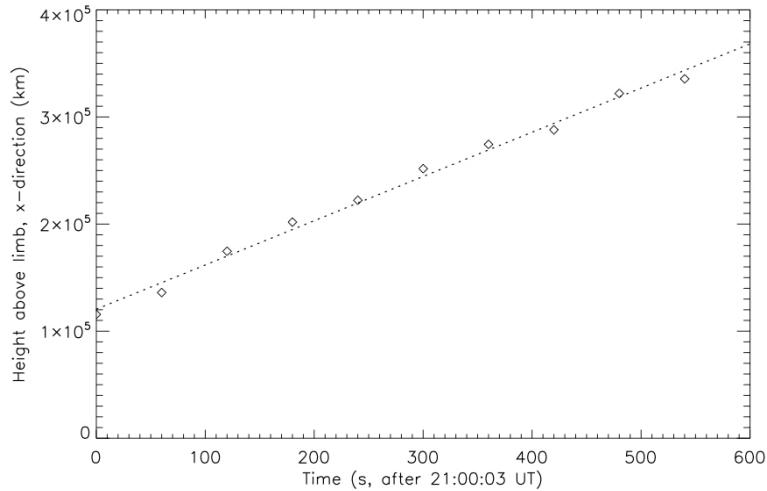

FIGURE 17c. A graph showing the height of the CME above the solar limb over time; the slope of the fitted line shows the average speed of the ejected mass to be 413 m s$^{-1}$.

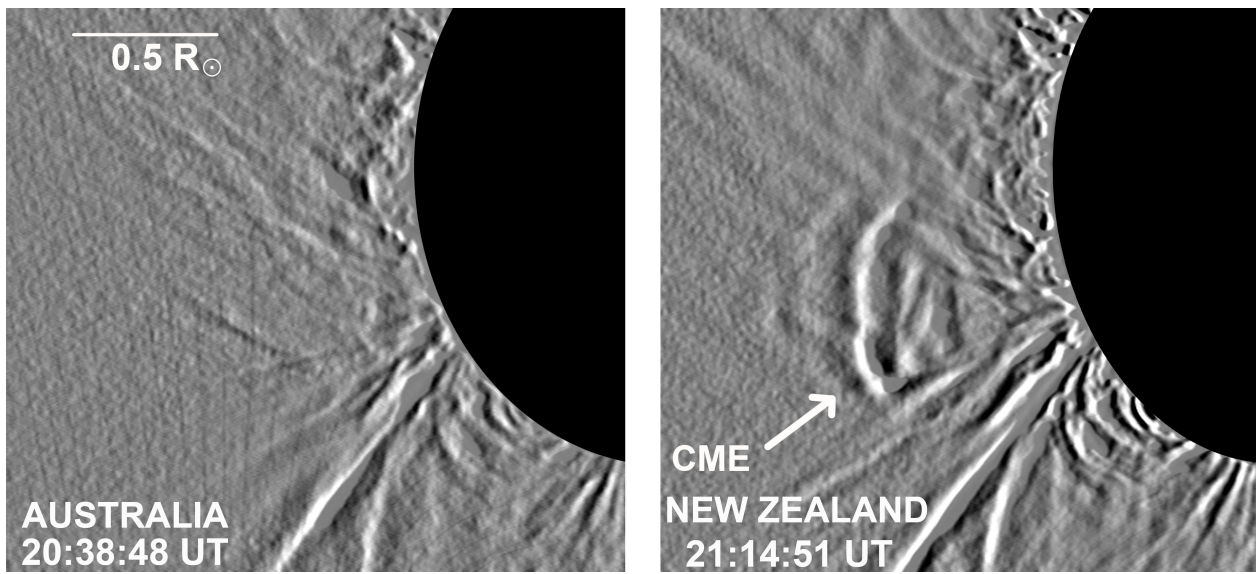

FIGURE 18a. Details of photographs obtained during the total solar eclipse from Mt. Carbine, Australia (left) and from a ship north of New Zealand (right). The time difference between the two photographs is 36 min. A CME is not detected in the Australia photograph but is immediately visible in the New Zealand photograph. The photographs were processed using the "Emboss" technique (Kim, et al., 2010). This filter enhances the edges and represents the rate of brightness change.

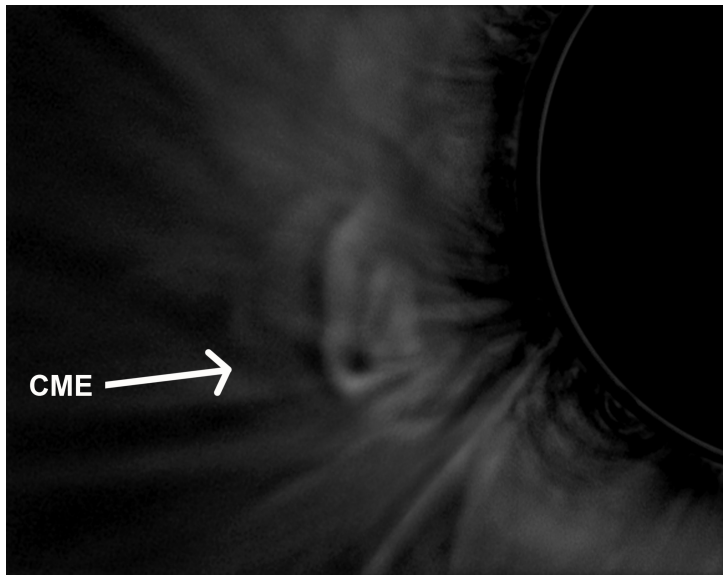

FIGURE 18b. The Australia photograph has been subtracted from the New Zealand photograph (see Figure 18a). Differences between the two photographs are, thus, enhanced. A CME is clearly visible, indicating increased activity in this region.

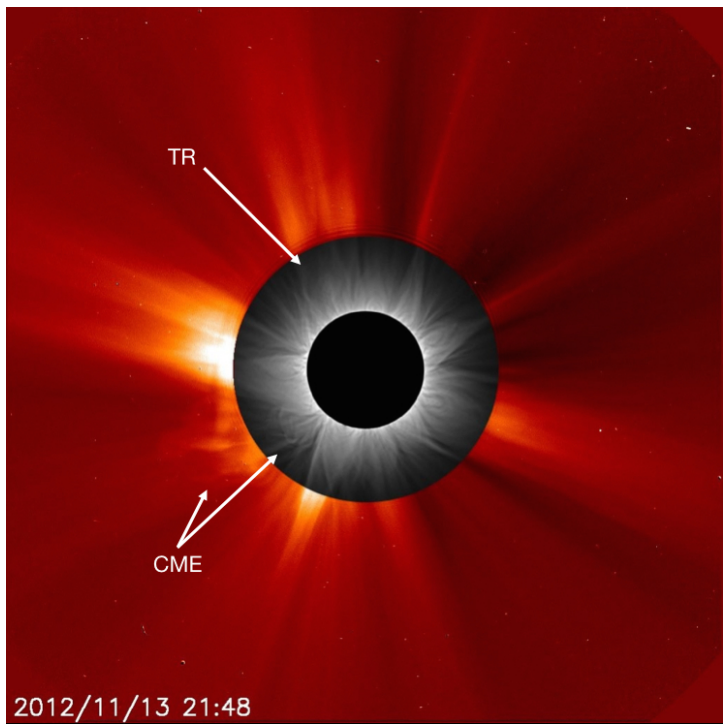

FIGURE 19. The *SOHO*/LASCO white-light corona image from the C2 coronagraph (UT 21:48) with our Figure-4-composite (UT 21:15) filling in the middle and inner corona; the same CME, temporally separated by 33 minutes, is labeled in the graph; the tennis-racquet-shaped structure (TR) is also visible. (Outer image courtesy of LASCO Consortium/NRL/NASA/ESA; inner image from NASA's Goddard Space Flight Center/NASA/ESA).

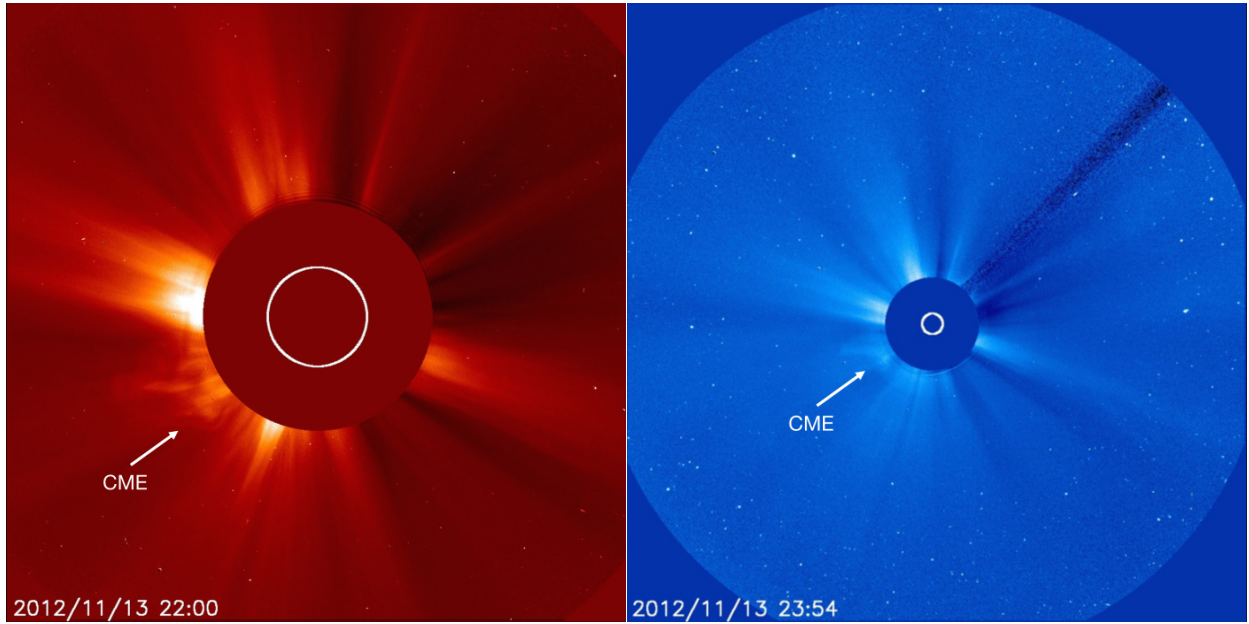

FIGURE 20. (a) The *SOHO*/LASCO white-light corona image from the C2 coronagraph, with the white circle marking the size and location of the solar photosphere. The selected LASCO images show the CME as it emerges from behind the occulting disk's lower left after the eclipse observations at 20:38 (Australia) and 21:15 (New Zealand ship). (Courtesy of LASCO Consortium/NRL/NASA/ESA) (b) A full *SOHO*/LASCO C3 image, illustrating the full range of the observed corona during the approximate time of the eclipse, with the CME in the lower left (courtesy of LASCO Consortium/NRL/NASA/ESA).

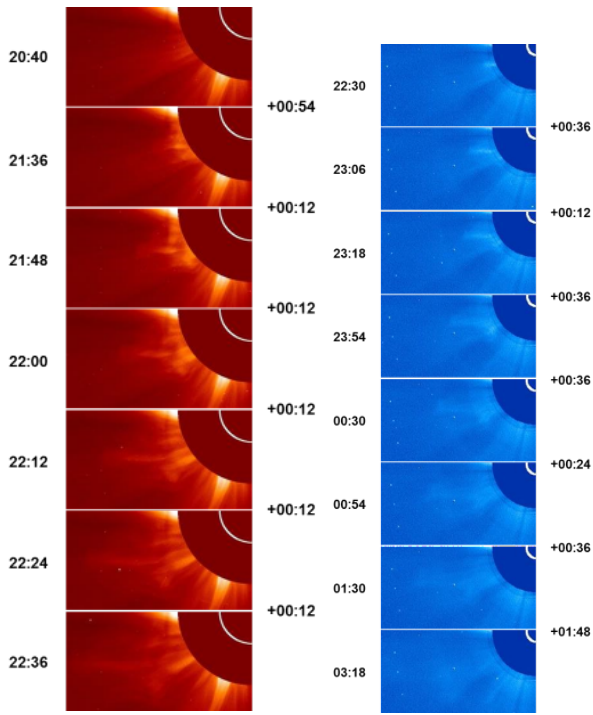

FIGURE 20. (c) Cut-outs of *SOHO*/LASCO images of the CME. Left: LASCO C2, which images from ~2–6 $R_\odot$; right: LASCO C3, with ~4 to ~15–20 $R_\odot$. (courtesy of LASCO Consortium/NRL/NASA/ESA).

## 6. DISCUSSION AND CONCLUSION

The composite eclipse images show the structure of the solar corona over the height range of 1–2 solar radii, largely unseen from spacecraft equipment, so they allow us to continuously connect coronal features over a wide range from the solar surface, seen in the UV, through the lower corona, and up to the outer corona seen with space coronagraphs, allowing us better to understand the development and location of dynamical structures on the Sun.

Our matched series of highly processed white-light eclipse images from the 2005, 2006, 2008, 2009, 2010, and 2012 total solar eclipses (Golub and Pasachoff, 2014) reveals the diminution of the solar-activity cycle through 2009, its modest resumption by the time of the 2010 total solar eclipse, and the onset of solar maximum by 2012. Morgan & Habbal (2010a) discuss the variations of the outer corona, as seen by LASCO, over the solar cycle. Since we see dynamical changes, we disagree with the conclusion of Woo (2010) that the variations seen in processed images result from differencing rather than actual coronal brightness.

Although activity on the Sun in the current cycle is very low when compared with the several previous cycles, the width of helmet streamers is nearly the same (Rušin et al., 2013). The different distributions of the white-light corona brightness and the emission in the EUV corona for the northern hemisphere illustrates that different mechanisms are responsible for the white-light corona and EUV emission corona.

Our composite images from *SDO*, EIT, or SWAP on the disk, through a doughnut of eclipse images, out through LASCO C2 and C3 allows features to be traced from their on-disk feet (at least for those with feet on the side of the Sun facing us, and the back side of the Sun can now be seen from *STEREO*) through the eclipse corona and into the outer corona. The magnetic-field structures (Figure 20) correspond well with the our processed eclipse images. *STEREO* views at that time (Figure 21) provides views of the CMEs from different angles.

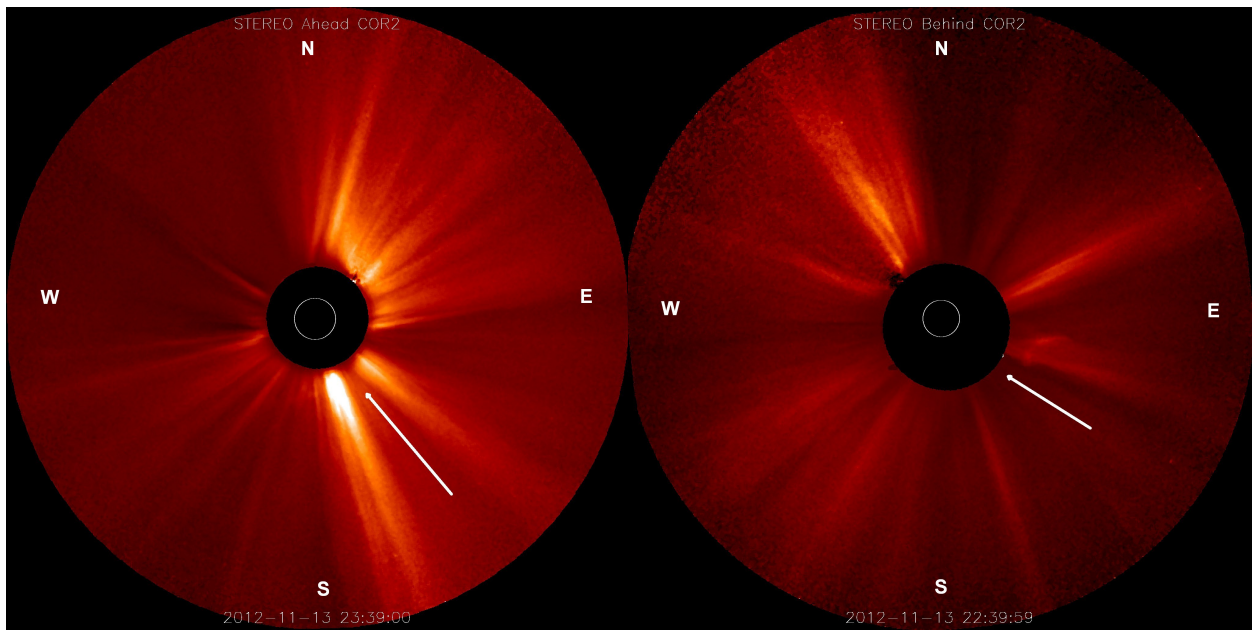

FIGURE 21. (a) Left: *STEREO*-A's view of the outer corona, with the CME visible as a faint arc slightly below the occulting disk. (b) Right: *STEREO*-B's view of the outer corona, with the CME visible as an arc to the right of the occulting disk. The arrows in both figures indicate the position of the observed CME as seen in Figures 17-20. The angle was determined with respect to the bright, south-east helmet streamer. STEREO A and B were separated by 108.6°, and were located at 124.0° and 127.4° from Earth, respectively (courtesy of *STEREO* Science Center).

As we discussed in Pasachoff et al. (2011a) and in Golub and Pasachoff (2010, 2014), the white-light eclipse images and the *SOHO*/LASCO images show photospheric light as scattered by coronal electrons held in place by unmeasurable coronal magnetic fields. The ultraviolet structures that we see on the solar disk with AIA, EIT, SWAP, or EUVE, on the other hand, show highly ionized coronal ions and so directly reveal the high-temperature coronal structures. Our high-resolution composited eclipse observations continue to show many fine-resolution rays similar to those discussed by Wang and Sheeley (2006) and Wang et al. (2007) from the 2006 eclipse.

Theories of coronal heating, including such alternatives as high-frequency waves or nanoflares (recently imaged by High Resolution Coronal Imager, Hi-C), have been summarized, among other places, by Golub & Pasachoff (2010); De Pontieu et al. (2011) have advanced a coronal-heating theory based on observations of Type II spicules and their energy inputs into the corona. Parnell and De Moortel (2012) have concluded that "the heating of the whole solar atmosphere must be studied as a highly coupled system," with "the dominant mechanisms still undetermined." Van Ballegooijen, et al. (2014) have evaluated images of footprints of coronal loops and the relation of their motion to coronal heating in active regions.

The resolution on CMEs in eclipse images is better than the resolution of the images from spacecraft, so total-eclipse observations remain interesting for comparison with the current generation of solar spacecraft. (The proposed PROBA3 spacecraft pair with an occulter hundreds of meters away from the telescope could eventually make images possible that cover the range we can now observe only with eclipses, but that eventuality is years away.) The availability of images from *STEREO*'s two spacecraft, from large angles around the Earth's orbit from Earth-orbiting spacecraft like *SDO*, allows for three-dimensional calculations of the angles traveled by CMEs (Mierla et al., 2008, 2010).

The eclipse's lunar occultation has an effect on Earth's atmosphere, tracked not only by changes in the clouds after first contact but also by a drop in temperature, which we normally measure but which was somewhat masked at this eclipse by the extensive cloud cover (see, for example, Peñaloza-Murillo & Pasachoff, 2013).

The recent observations of the 2013 total solar eclipse from Africa (Pasachoff et al., 2014ab) will be linked with observations of the 2015 total solar eclipse from the Arctic, including Svalbard, and the 2016 total solar eclipse from Indonesia and the Pacific, as part of the run-up to the 2017 total solar eclipse that will cross the continental U.S. from Oregon to South Carolina for not only scientific considerations but also for public outreach (Hudson et al., 2011; Habbal et al. 2011b, Pasachoff, 2014).

## ACKNOWLEDGMENTS

J.M.P.'s research on the annular and total solar eclipses of 2012 is supported in part by the Solar-Terrestrial Program of the Atmospheric and Geospace Sciences Division of the National Science Foundation through grant AGS–1047726; the 2012 expedition received additional support from the Brandi Fund, the Rob Spring Fund, the Milham Meteorology Fund, and Science Center funds from Williams College. His current eclipse work, for the 2015 total solar eclipse, is supported by grant 9616-14 from the Committee for Research and Exploration of the National Geographic Society. The work of V.R. and M.S. was partially supported by the VEGA grant agency projects 2/0098/10 and 2/0003/13 (Slovak Academy of Sciences) and by NGS Grant 0139–12. For the Australian expedition, we thank Nikon Professional Services and Williams College's Equipment Loan Office/James Lillie for providing equipment. We thank Wendy Carlos for her excellent computer work in providing a composite image of the overall coronal structure based on our imaging. We are very grateful to Kasuo Shiota for sharing his excellent observations with us. J.M.P. thanks Andy Ingersoll and the Planetary Sciences Department of the California Institute of Technology for hospitality during the period we composited images and completed the revision. Nicholas Weber of the Dexter Southfield School, Brookline, Massachusetts, was a full participant with the RED Epic cameras and other equipment on site. We also thank Alec Engell of Montana State University and Robert Lucas of the University of Sydney for their helpful participation on site. The assistance of Terry Cuttle of the Queensland Amateur Astronomers was invaluable for finding our sites. For the loan of tracking mounts, we thank Drs. Joe Brimacombe and Tim Carruthers of Cairns and Charles Frank of Adelaide. Support for D.B.S. and SWAP came from PRODEX grant no. C90345 managed by the European Space Agency in collaboration with the Belgian Federal Science Policy Office (BELSPO) in support of the *PROBA2*/SWAP mission, and from the European Commission's Seventh Framework Programme (FP7/ 2007–2013) under the grant agreement no. 218816 (SOTERIA project, www.soteria-space.eu). SWAP is a project of the Centre Spatial de Liège and the Royal Observatory of Belgium funded by the Belgian Federal Science Policy Office (BELSPO). We are grateful for the assistance of David Rust (JHU/APL). We thank Todd Hoeksema of Stanford Solar Observatory for providing the Carrington-cycle magnetic-field map. Finally, we are very grateful to Zuzana Kanuchova in Slovakia for help with obtaining and completing the final form of some of the figures.

Facilities: *PROBA2*, *SDO*, *SOHO*, *STEREO*, *Wilcox Solar Observatory*

Supplemental Material:. Additional *SDO* images at UTC 2012/11/13, taken near to totality at the Queensland eclipse. Comparison shows different levels of solar activity. *SDO*/AIA filters at white light (photospheric continuum, log T = 3.7), 171 Å (Fe IX, 5.8), 335 Å (Fe XVI, 6.4), 1600 Å (C IV + continuum, 5.0), 1700 Å (continuum, 3.7), 94 Å (Fe XVIII, 6.8), 304 Å (He II, 4.7), 131 Å (Fe XIII, XX, XXIII; 5.6, 7.0, 7.2), and 211 Å (Fe XIV, 6.3). *SDO*/AIA takes a full set of 8 filtered images, in two groups of 4, with an approximate cadence of 12 s. See Lemen et al. (2011). (Courtesy: *SDO*/LMSAL/SAO/NASA)